\documentclass[prd,twocolumn,aps,tightenlines,floatfix,showpacs,byrevtex,nofootinbib,preprintnumbers]{revtex4}
\usepackage{epsfig}
\usepackage{dcolumn}
\usepackage{amsmath,amssymb,graphicx,makeidx}

\def\be{\begin{equation}}
\def\ee{\end{equation}}

\begin{document}


\preprint{ADP-03-109/T547}

\title{Delta Baryon Magnetic Moments From Lattice QCD}
\author{ I.C. Cloet\footnote{icloet@physics.adelaide.edu.au},  
D.B. Leinweber\footnote{dleinweb@physics.adelaide.edu.au}  and
A.W. Thomas\footnote{athomas@physics.adelaide.edu.au}}

\affiliation{Special Research Centre for the Subatomic Structure of 
             Matter and      \\
             Department of Physics and Mathematical Physics, University of Adelaide,
             SA 5005, Australia}

\begin{abstract}
Theoretical predictions for the magnetic moments of the physical 
${\Delta}$ baryons are extracted from lattice   
QCD calculations. We utilize finite-range regulated effective
field theory that is constructed
to have the correct Dirac moment mass dependence in the region where the {\em u} and 
{\em d} quark masses are heavy.
Of particular interest is the chiral nonanalytic behaviour encountered as the $N\, \pi$
decay channel opens.
We find  a $\Delta^{++}$ magnetic moment (at the $\Delta$ pole) of 
$\mu_{\Delta^{++}}=4.99 \pm 0.56\ \mu_N$. This result is 
within the Particle Data Group range of 3.7--7.5~$\mu_N$  
and compares well with the experimental result of Bosshard {\em et al.} of 
$\mu_{\Delta^{++}}=4.52 \pm 0.51 \pm 0.45\ \mu_N$. The interplay between the different
pion-loop contributions to the $\Delta^+$ magnetic moment leads to the surprising
result that the proton moment may exceed that of the $\Delta^+$, 
contrary to conventional expectations.      
\end{abstract}


\maketitle

\newpage

\begin{section}{Introduction}

The magnetic moments of the $\Delta$ baryons have already caught the attention
of the experimental community and hold the promise 
of being accurately measured in the 
foreseeable future. Experimental estimates exist for 
the ${\Delta^{++}}$, based on the reaction $\pi^+\ p\ \to \pi^+\ \gamma'\ p$.  
The Particle Data Group \cite{Hagiwara:pw} cites a range of values, 
3.7--7.5~$\mu_N$,  
for the ${\Delta^{++}}$ magnetic moment, with the two most recent 
experimental results being $\mu_{\Delta^{++}}=4.52 \pm 0.51 \pm 0.45\ \mu_N$ 
\cite{Bosshard} 
and $\mu_{\Delta^{++}}=6.14 \pm 0.51\ \mu_N$ \cite{LopezCastro:2000ep}.
This gives an idea of the elusive nature of the $\Delta$ magnetic moments.
In principle, the $\Delta^+$ magnetic moment can be obtained from the reaction 
$\gamma\ p\ \to \pi^o\ \gamma'\ p$, as demonstrated at the Mainz microtron 
\cite{Kotulla,Drechsel:2001qu} and  Kotulla {\it et al.} have recently
 reported an initial measurement of $\mu_{\Delta^+} = 2.7^{+1.0}_{-1.3} {\rm (stat.)}
 \pm 1.5 {\rm (syst.)} \pm 3 {\rm (theor.)}~\mu_N$ \cite{Kotulla:2002cg}.

Recent extrapolations of octet baryon magnetic moments have utilized 
a Pad$\acute{\rm e}$ approximant 
\cite{Leinweber:1998ej,Hackett-Jones:2000qk,Leinweber:2001ui,Cloet:2002eg}, 
an analytic continuation of chiral perturbation theory ($\chi$PT) which 
incorporates the correct leading nonanalytic (LNA) 
structure of $\chi$PT. 
This idea was recently extended to a study of decuplet baryons in the Access Quark Model
\cite{Cloet}, where the next-to-leading nonanalytic (NLNA) structure of
$\chi$PT was also included. Incorporating the NLNA terms into 
the extrapolation function contributes little to the octet baryon magnetic 
moments, however it proves vital for the decuplet. This is because the NLNA 
terms from $\chi$PT contain information regarding the branch point at 
the octet-decuplet mass-splitting, $m_{\pi} = \delta$, associated 
with the $\Delta \to N \pi$ decay channel, which plays a significant role in 
decuplet magnetic moments. Here we include the NLNA behaviour in a chiral
extrapolation of the magnetic moments of the $\Delta$ baryons calculated in lattice QCD.

Although the Pad$\acute{\rm e}$ technique provides a well behaved 
method of extending 
$\chi$PT to heavy quark masses, it is not so obvious how it could be extended
to higher order in a chiral expansion. We have therefore chosen to 
calculate explicitly the pion loop diagrams which give rise to the LNA and the 
NLNA behaviour of the $\Delta$ magnetic moments using finite-range regularization
\cite{Young:2002ib,Donoghue:1998bs}. In this investigation we select the 
dipole-vertex regulator.
We stress that, in the context of the chiral extrapolation of lattice QCD data,
the use of a finite-range regulator is the preferred alternative to dimensional 
regularization as demonstrated in Ref.~\cite{Young:2002ib}.
Ideally, the regulator mass should be constrained by lattice QCD results.
However, the data available lies at large quark masses and provides little guidance.
As such, we explicitly explore the  
regulator-mass dependence of our results, finding a sensitivity of 
3.6\% or less in the charged $\Delta$ magnetic moments, for variation of the dipole mass 
parameter in the range 0.5--1.0\ GeV. 

\end{section}


\begin{section}{Extrapolations -- An Effective Field Theory}

To one-loop order the dimensionally-regulated $\chi$PT 
result for decuplet baryon magnetic moments is \cite{Banerjee:1995wz,Correction} 

\begin{multline}
\mu_i = a_0 + \sum_{j= \pi, K} \frac{M_N}{32 {\pi}^2 f_j^2} 
      \Bigl( \alpha_j^i\, \frac{4}{9}\, {\cal H}^2\, {\cal F}(0,m_j,\mu) \\ 
           + \beta_j^i\, {\cal C}^2\, {\cal F}(-\delta,m_j,\mu) \Bigr)
           + a_2\, m_{\pi}^2\, .
\label{CPT}
\end{multline}

\noindent The ${\cal F}(\delta,m_j,\mu)$ functions 
(with $\delta$ the octet-decuplet mass difference)
are the nonanalytic 
components of the meson loop diagrams shown in Fig.~1. These expressions are 
given explicitly in Ref.~\cite{Banerjee:1995wz}. The constants $a_0$ and $a_2$ 
multiplying the analytic terms are 
not specified within $\chi$PT and must be determined by other means. 

It is well known that dimensionally-regulated $\chi$PT expansions for 
baryon properties are troubled by a lack of clear 
convergence \cite{Hatsuda:tt,SVW,Bernard:2002yk,Thomas:2002sj}. 
This problem must be addressed if $\chi$PT results are to be applied
far beyond the chiral limit, for example in the extrapolation of lattice 
data where pion masses up to 1 GeV are involved. Indeed, truncated 
dimensionally-regulated $\chi$PT expansions
for magnetic moments exhibit the incorrect behaviour for large $m_{\pi}$.
To address these concerns we adopt finite-range regulated $\chi$PT by
calculating the meson-loop 
diagrams using a dipole-vertex regulator.  The loop contributions 
then approach zero naturally as $m_{\pi}$ becomes large.
We combine the analytic terms of the chiral expansion 
into a single term that maintains the chiral expansion to the order we are working, 
while guaranteeing the correct magnetic moment
mass dependence at heavy pion masses. 
%
\begin{figure}[tbp]
\begin{center}
{\epsfig{file=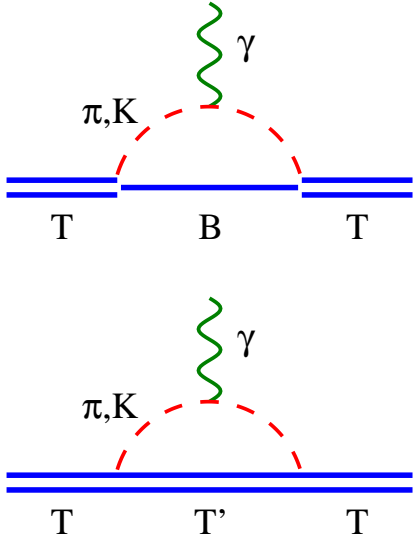, height=8cm, angle=0}}
\caption{Diagrams indicating the leading and next-to-leading contributions included in the 
calculations. The letters 
T and B correspond to  decuplet and octet baryons respectively. 
The dashed line is the appropriate
meson accompanying the baryon transition. 
The analytic expressions for these diagrams are given 
explicitly in Eqs.~(\ref{loops}).}                 
\label{diagrams}
\end{center}
\end{figure}

The extrapolation function which we employ to obtain decuplet baryon 
magnetic moment predictions from lattice QCD data, thus takes the form
\begin{multline}
\mu = \frac{a_0}{1+ c_0 \, m_{\pi}^2} + 
\sum_{T'} \chi^{j}_{T'}~ G_{TT'}(m_j, \Lambda) \\    
+ \sum_{B} \chi^{j}_{B}~ G_{TB}(m_j, \delta, \Lambda)\, ,
\label{MM}
\end{multline}
where the sums over $T'$ and $B$ include the decuplet and octet 
intermediate states, respectively. 
The first term encapsulates the pion mass dependence of the 
photon field coupling
directly to the baryon and reflects previous success 
in the extrapolation of octet baryon 
magnetic moments \cite{Leinweber:1998ej,Hackett-Jones:2000qk,Cloet:2002eg}. 
This term ensures the correct behaviour, as a function of quark mass,  
of the Dirac moments of the quarks  
at heavy quark masses (as $m_{\pi}^2 \propto m_q$ up to 1 GeV$^2$ -- above 
this range one should use $m_q$ directly). 
The parameters $a_0$ and $c_0$ are chosen to optimize the  
fit to lattice data. The functions 
$G_{TT'}(m_j, \Lambda)$ and 
$G_{TB}(m_j, \delta, \Lambda)$ are the (heavy baryon) loop contributions to the 
magnetic form factors, in the limit $q \to 0$, for 
decuplet-decuplet and decuplet-octet 
transitions respectively, see Fig.~1. These functions are given by
\begin{align}
G_{TT'}(m_j, \Lambda) &= \lim_{q \to 0}\ 
\frac{1}{2\pi} \int\ {\rm d}^3 k\ {\cal U}(k)\ {\cal U}(k')\
  \frac{(\hat{q}\times \vec{k})^2}{(\omega_k \ \omega_{k'})^2}\, , \nonumber \\
G_{TB}(m_j, \delta, \Lambda) &=
\lim_{q \to 0}\ 
\frac{1}{2\pi} \int\ {\rm d}^3 k\ {\cal U}(k)\ {\cal U}(k')  \nonumber \\
& \hspace{10mm} \frac{(\omega_k + \omega_{k'} - \delta)(\hat{q}\times \vec{k})^2 }
{\omega_k\ \omega_{k'}(\omega_k + \omega_{k'})(\omega_k - \delta)
(\omega_{k'} - \delta)}\, , 
\label{loops}
\end{align} 

\noindent where  $q,~k,~k'$ are the momenta of the photon, incoming meson and outgoing meson, 
respectively. Note, $\vec{k}' = \vec{k} + \vec{q}$ and $\omega_k=\sqrt{k^2+m_j^2}$. 
The functions ${\cal U}(k)$ and  ${\cal U}(k')$ 
are used to regulate the loop integrals - we use a dipole with a finite 
mass parameter $\Lambda$. Therefore

\begin{align}
{\cal U}(k)  &= \left(\frac{\Lambda^{2}}{\Lambda^2+k^2} \right)^2\, , \\
{\cal U}(k') &= \left(\frac{\Lambda^{2}}{\Lambda^2+(k')^2} \right)^2\, .
\end{align}
\begin{table}[tbp]
\begin{center}
\addtolength{\tabcolsep}{-2.2pt}
\begin{ruledtabular}
\begin{tabular}{lclclclcc}

              \multicolumn{2}{c}{\emph{$\Delta^{++}$}}
             &\multicolumn{2}{c}{\emph{$\Delta^{+}$}}
             &\multicolumn{2}{c}{\emph{$\Delta^{0}$}}
             &\multicolumn{2}{c}{\emph{$\Delta^{-}$}}    \\[+0.4ex]

Channel         & $\chi$~~   & Channel          & $\chi$~~  & Channel         & $\chi$~~   & Channel       & $\chi$~~   \\[+0.5ex] \hline \\[-1.8ex]

~$\Delta^+$~    & 0.265~   & ~$\Delta^0$~     & ~0.353~ & ~$\Delta^{-}$~  & ~0.265~  & ~$\Delta^0$~  & -0.265~ \\
~$\Sigma^{*+}$~ & 0.184~   & ~$\Delta^{++}$~  & -0.265~ & ~$\Delta^{+}$~  & -0.353~  & ~$n$~         & -0.795~ \\
~$p$~           & 0.795~   & ~$\Sigma^{*0}$~  & ~0.123~ & ~$\Sigma^{*-}$~ & ~0.061~  &               &         \\
~$\Sigma^{+}$~  & 0.552~   & ~$n$~            & ~0.265~ & ~$p$~           & -0.265~  &               &         \\
                &          & ~$\Sigma^{0}$~   & ~0.368~ & ~$\Sigma^{-}$~  & ~0.184~  &               &         \\
\end{tabular}
\end{ruledtabular}
\end{center}
\caption{The chiral coefficients for each intermediate state for $\Delta$ baryon transitions.}        
\label{table:ChiralC5}
\end{table}

The masses, $m_j$, of Eq.~(\ref{MM}) are octet-meson masses associated 
with the meson cloud
and $\chi_{i}$ are model independent constants \cite{Li:1971vr} that give the 
magnitude of the loop contribution
for each intermediate baryon state. These coefficients are given by \cite{Banerjee:1995wz}
\begin{eqnarray}
\chi^{(\pi)}_{T_i} = \frac{M_N\ {\cal H}^2}{72\, {\pi}^2 \, 
(f_{\pi})^2}\ {\alpha^{(\pi)}_{T_i}}\, , \hspace{2.5mm}    
\chi^{(K)}_{T_i}   = \frac{M_N\ {\cal H}^2}{72\, {\pi}^2\, 
(f_{K})^2}\ {\alpha^{(K)}_{T_i}}\, ,                      \nonumber \\ 
\chi^{(\pi)}_{B_i} = \frac{M_N\ {\cal C}^2}{32\, {\pi}^2\, 
(f_{\pi})^2}\ {\beta^{(\pi)}_{B_i}}\, , \hspace{3.0mm} 
\chi^{(K)}_{B_i}    = \frac{M_N\ {\cal C}^2}{32\, {\pi}^2\, 
(f_{K})^2}\ {\beta^{(K)}_{B_i}}\, ,     
\label{CC}
\end{eqnarray}
where $\cal H$ describes meson couplings to decuplet 
baryons and $\cal C$ is the 
octet--decuplet coupling constant. We assign ${\cal H}$ and  
${\cal C}$ their $SU(6)$ values of ${\cal H} = -3 D$ and 
${\cal C} = -2 D$ \cite{Butler:1992pn} where the tree level value for $D$ is $0.76$.
The decay constants take the values $f_{\pi}=93~\rm MeV$ and $f_{K}= 1.2\, f_{\pi}\rm$ 
\cite{Jenkins:1992pi}, appropriate to an expansion about the chiral SU(2) limit.
Note also that $M_N$ is the
nucleon mass and the parameters $\alpha$ and $\beta$ are given in Ref.~\cite{Butler}.
The model independent loop coefficients, $\chi$, are summarized in 
Table~\ref{table:ChiralC5}.

In these calculations the complex mass scheme \cite{LopezCastro:2000ep,ElAmiri:xa} is adopted as a 
method of incorporating the finite life of the $\Delta$ resonances, 
whilst also retaining electromagnetic gauge invariance. That is, one aims to extract the 
magnetic moment of 
the $\Delta$ at the pole position in the complex energy plane, 
$\delta^{\rm (pole)} \equiv \delta_R - i\, \delta_I = 270 - i\, 50$~MeV \cite{Hagiwara:pw} 
for the physical pion mass ($m_{\pi}^{\rm phys}$).
Since the value at the position of the $\Delta$ pole is independent of the path chosen,
we illustrate extrapolations along the path $\delta_R$ constant and $\delta_I$ given by
\begin{equation}
\delta_{I} =
 G\ \pi\ (\delta_R^2 - m_{\pi}^2)^{\frac{3}{2}}
\left(\frac{\Lambda^2}{\Lambda^2 + \delta_R^2- m_{\pi}^2} \right)^4\ \Theta(\delta_R - m_{\pi})\, . 
\label{complex}
\end{equation}
The latter is motivated by the usual expression for the $\Delta \to N\, \pi$ 
self-energy \cite{Young:2002ib}, with $G$ chosen to ensure $\delta_I = 50$~MeV at $m_{\pi}^{\rm phys}$.

To relate the kaon and pion masses  we utilize the following relations provided by $\chi$PT 
\begin{align}
m_K^2   &= {m_K^{(0)}}^2 + \frac{1}{2} m_{\pi}^2\, ,  \\
m_K^{(0)} &= \sqrt{(m_K^{\mathrm{phys}})^2 - 
\frac{1}{2}(m_{\pi}^{\mathrm{phys}})^2}\, .
\end{align}
This allows us to also
incorporate kaon loops (at fixed strange quark mass) 
in extrapolations as a function of $m_{\pi}^2$,
to the physical mass regime.
%
\begin{figure}[tbp]
\begin{center}
{\epsfig{file=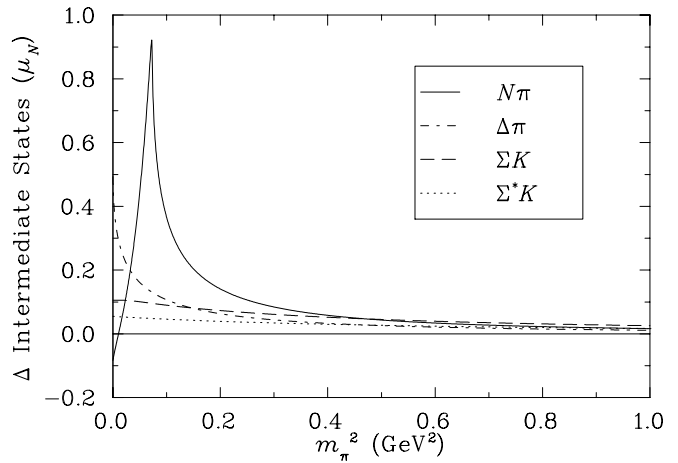, height=\hsize, angle=90}}
\caption{Plots of the four types of loop contributions to the $\Delta$ magnetic moments 
(without the coefficients) that
are considered here. These plots are obtained from Eqs.~(\ref{loops}),
where we have used the complex mass scheme as a gauge 
invariant method with which to incorporate the finite life-time of the $\Delta$ resonances.}          
\label{fig:fun1}
\end{center}
\end{figure}
\begin{figure}[tbp]
\begin{center}
{\epsfig{file=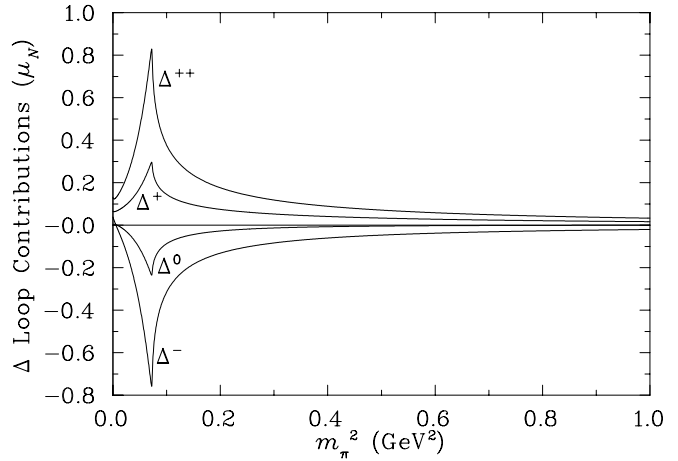, height=\hsize, angle=90}}
\caption{Plots of the total loop contribution 
considered for each $\Delta$ baryon.  
These curves are the sum of the functions plotted in Fig.~2 
multiplied by the appropriate coefficients
listed in Table~\ref{table:ChiralC5}.}        
\label{fig:fun2}
\end{center}
\end{figure}

If we expand the first term of Eq.~(\ref{MM}) as a Taylor 
series about $m_{\pi} = 0$ to order 
${\cal O}(m_{\pi}^4)$,
and take the limit as $\Lambda \to \infty$ in the loop integrals,  
we obtain the traditional $\chi$PT expansion for 
decuplet magnetic moments, as given in Eq.~(\ref{CPT}) 
(where the infinite constants encountered in dimensional regularization simply redefine  
$a_0$). Further, since the loop contributions approach 
zero much faster than $1/m_{\pi}^2$ for any 
reasonable value of $\Lambda$, Eq.~(\ref{MM}) guarantees the correct  
mass dependence of the Dirac magnetic moment 
in the heavy quark mass regime. From previous studies of the 
nucleon axial form factor \cite{Guichon:1982zk}
it has been consistently demonstrated that $\Lambda$ in the dipole 
regulator must have a magnitude $<$ 1 GeV. For this 
investigation we assign $\Lambda = 0.8$ GeV, the optimal value obtained in 
analyses of state of the art full QCD simulations of the nucleon mass
\cite{Young:2002ib}.

In Fig.~2 we plot 
Eqs.~(\ref{loops}) (without the chiral coefficient prefactors) with the above value of $\delta$ 
and a regulator of $\Lambda=0.8$ GeV.
Four types of intermediate baryon states are considered.
The opening of the $N \, \pi$ decay channel has an interesting effect on
the magnetic moment contribution.  
Fig.~3 presents a plot of the total loop contribution for each $\Delta$ 
baryon, determined using Eqs.~(\ref{loops}) and the chiral coefficients given in Table~\ref{table:ChiralC5}.

\end{section}


\begin{section}{Results}

\begin{table*}[t]
\begin{center}
\begin{ruledtabular}
\begin{tabular}{ccccccc}
kappa    &Baryon Mass   &Pion Mass    &$\Delta^{++}$  &$\Delta^+$    &$\Delta^-$     &$\Delta^-$\\
         & (GeV)        & (GeV)       & ($\mu_N$)     & ($\mu_N$)    & ($\mu_N$)     & ($\mu_N$)\\[+0.7ex] 
\hline \\[-1.7ex]
0.152    &1.74~(50)     & 0.964~(12)  &2.81~(18)      &1.40~~(9)     &0.000~(00)     &-1.40~~(9)\\
0.154    &1.57~(60)     & 0.820~(11)  &3.19~(28)      &1.59~(14)     &0.000~(00)     &-1.59~(14)\\
0.156    &1.39~(80)     & 0.665~(13)  &3.67~(43)      &1.83~(21)     &0.000~(00)     &-1.83~(21)\\
\end{tabular}
\end{ruledtabular}
\end{center}
\caption{Quenched lattice QCD magnetic moments for the $\Delta$ baryons 
extracted from Ref.~\protect\cite{Leinweber:1992hy}. Uncertainties are obtained using a
third-order single-elimination jack-knife error analysis. 
Lattice results for the $\Delta^0$ are exactly zero, in the quenched approximation.}   
\label{table:Ldata}
\end{table*}

The lattice QCD results given in Table~\ref{table:Ldata} 
are extracted from Ref.~\cite{Leinweber:1992hy}. 
These lattice calculations employ sequential-source three-point function 
based techniques \cite{Comment} utilizing the conserved 
vector current such that no renormalization 
is required in relating the lattice results to the continuum. These simulations
utilized twenty-eight quenched gauge configurations on a $24 \times 12 \times
12 \times 24$ periodic lattice at $\beta = 5.9$, corresponding to a lattice 
spacing of 0.128(10)~fm. Moments are obtained from 
the form factors at $0.16~\rm GeV^2$ 
by assuming equivalent $q^2$ dependence for both the 
electric and magnetic form factors. This assumption will be tested in future
studies incorporating the ideas presented here.
Uncertainties are statistical in origin and are estimated by a 
third-order single-elimination jack-knife analysis \cite{BE}. 

In Figs.~4--7 we present fits of the extrapolation function, Eq.~(\ref{MM}),
to the lattice data as a function of $m_{\pi}^2$.   The resulting magnetic moment predictions, along 
with the fit parameters $a_0$ and $c_0$, are 
summarized in Table~\ref{table:Results}.  
In Fig.~4 we show two experimental values for 
the $\Delta^{++}$ magnetic moment, the 
result $\mu_{\Delta^{++}}=4.52 \pm 0.50 \pm 0.45\ \mu_N$ 
is from Ref.~\cite{Bosshard} 
and $\mu_{\Delta^{++}}=6.14 \pm 0.51\ \mu_N$ is given 
in  Ref.~\cite{LopezCastro:2000ep}.
The discrepancy between these two results is a reasonable 
indication of the current level of systematic error
in the experimental determination of the $\Delta^{++}$ magnetic 
moment. Our prediction of 
$\mu_{\Delta^{++}}=4.99 \pm 0.56\ \mu_N$ agrees 
well with the first experimental result, however it lies slightly below 
the range of the second. We note that the approach 
used in Ref.~\cite{LopezCastro:2000ep} does not respect unitarity and 
when this shortcoming
is addressed the authors find  $\mu_{\Delta^{++}}=6.01 \pm 0.61\ \mu_N$ \cite{LopezCastro:2000ep},  
resulting in a somewhat better 
agreement between our prediction and this experimental 
result.  The result reported here for the $\Delta^+$ magnetic moment of $2.49 \pm 0.27\ \mu_N$ 
is in agreement with the
initial measurement of Kotulla {\it et al.} \cite{Kotulla:2002cg}, namely 
$\mu_{\Delta^+} = 2.7^{+1.0}_{-1.3} {\rm (stat.)} \pm 1.5 {\rm (syst.)} \pm
3 {\rm (theor.)} \mu_N$, however the experimental error at this time is
still very large. 

To address the issue of regulator-mass dependence we vary 
$\Lambda$ between 0.5 and 1.0~GeV, in each case readjusting $a_0$ and $c_0$ to 
fit the lattice data and find only a slight regulator-mass dependence 
of 3.6\% or less, for each of the charged
$\Delta$ baryons. 
Ultimately, lattice results at lighter quark masses will constrain $\Lambda$
and therefore reduce this uncertainty.

The inclusion of  
the $\Delta^0$ is more for completeness rather
than a firm result, as the lattice data in 
quenched QCD yields identically zero in this case.
By varying the regulator
$\Lambda$ between 0.5 and 1.0~GeV we find the moment remains positive,
with the order of magnitude of our result unchanged. 
We await new, unquenched lattice 
data in order to obtain a stronger prediction for the 
$\Delta^0$ magnetic moment. 
%
\begin{figure}[tbp]
\begin{center}
{\epsfig{file=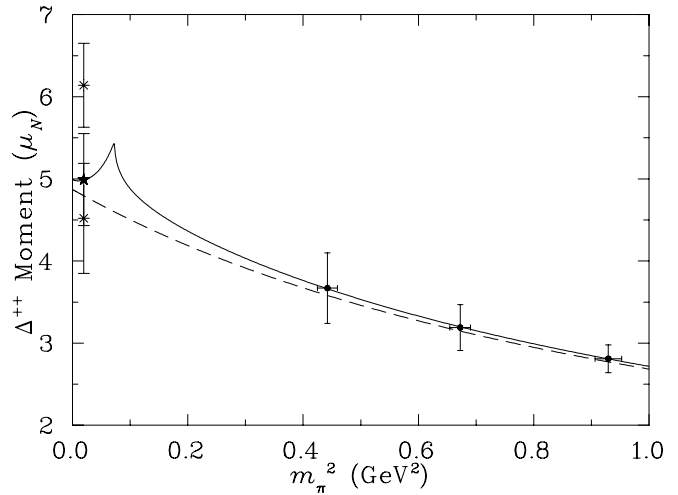, height=\hsize, angle=90}}
\caption{The fit (solid line) to the lattice QCD data (solid circles) for the $\Delta^{++ }$ 
magnetic moment using the extrapolation function, 
Eq.~(\protect\ref{MM}).  The dashed line is the plot 
of the first term of Eq.~(\protect\ref{MM}) -  and reflects the 
analytic terms of the chiral expansion.
The solid star ($\star$) indicates the prediction of the extrapolated 
lattice data.
The two most recent experimental results are 
indicated by the two asterisks ($\ast$), where 
the smaller prediction is 
taken from Ref.~\protect\cite{Bosshard} and 
the larger from  Ref.~\protect\cite{LopezCastro:2000ep}.}              
\label{fig:d++}
\end{center}
\end{figure}

\begin{table*}[tbp]
\begin{center}
\begin{ruledtabular}
\begin{tabular}{lcccc}
Baryon          &$a_0$   &$c_0$  &Lattice~($\mu_N$)  & Experiment~($\mu_N$) \\
\hline \\ [-2.0ex]
$\Delta^{++}$   &~4.87    &0.82   &~4.99~(56)          & 3.7--7.5   \\
$\Delta^{+}$    &~2.44    &0.83   &~2.49~(27) & $2.7^{+1.0}_{-1.3} {\rm (stat.)} \pm 1.5 {\rm (syst.)} \pm 3 {\rm (theor.)}$ \\
$\Delta^{0}$    &~0.63    &381    &~0.06~(00)          &            \\
$\Delta^{-}$    &-2.40    &0.80   &-2.45~(27)          &
\end{tabular}
\end{ruledtabular}
\end{center}
\caption{Theoretical predictions for the four $\Delta$ baryon 
magnetic moments from extrapolated 
lattice QCD results. The quoted errors are statistical in origin.
The fit parameters $a_0$ and $c_0$ are given for each scenario. 
The experimental 
values for the $\Delta^{++}$ and $\Delta^{+}$ magnetic moments are taken from
Ref.~\protect\cite{Hagiwara:pw}. and Ref.~\protect\cite{Kotulla:2002cg}, respectively.}    
\label{table:Results}
\end{table*}

The interesting feature of the plots given in Figs.~4--7 is the cusp at $m_{\pi} = \delta$ 
which indicates the opening of the octet decay channel, $\Delta \to N \pi$. 
The physics behind the cusp is intuitively revealed by the 
relation between the 
derivative with respect to $m_{\pi}^2$ of the magnetic moment 
and the derivative
with respect to the momentum transfer, $q^2$, provided by the pion propagator
$1/(q^2+m_{\pi}^2)$ in the heavy baryon limit. 
Derivatives with respect to $q^2$
are proportional to the magnetic radius in the limit $q^2 \to 0$ 
\begin{equation}
\langle r_M^2 \rangle = \left. -6\, \frac{d\, G_M(q^2)} {dq^2}\right|_{q^2=0}\, .
\end{equation}  
If we consider for example $\Delta^{++} \to p \pi^+$,  
with $|j,m_j\rangle = |3/2,3/2\rangle$,  
the $N\, \pi$ state is in  relative P-wave orbital 
angular momentum with 
$|l,m_l\rangle = |1,1\rangle$. Thus the pion makes a positive 
contribution to the magnetic moment.
As the opening of the $p\, \pi^+$ decay channel is approached from 
the heavy quark-mass regime, 
the range of the pion cloud increases. Just above threshold the 
pion cloud extends towards 
infinity as $\Delta E \to 0$, $\Delta E \Delta t \sim \hbar$ 
and the radius of the magnetic form factor  
diverges similarly, $ ({\partial}/{\partial q^2}) G_M \propto ({\partial}/{\partial m_{\pi}^2}) 
G_M \to - \infty$ from above threshold.
Below threshold, $G_M$ becomes complex and the magnetic moment 
of the $\Delta$ is identified 
with the real part. The imaginary part describes the physics 
associated with photon-pion 
coupling in which the pion is subsequently observed as a decay product. 
%
\begin{figure}[tbp]
\begin{center}
{\epsfig{file=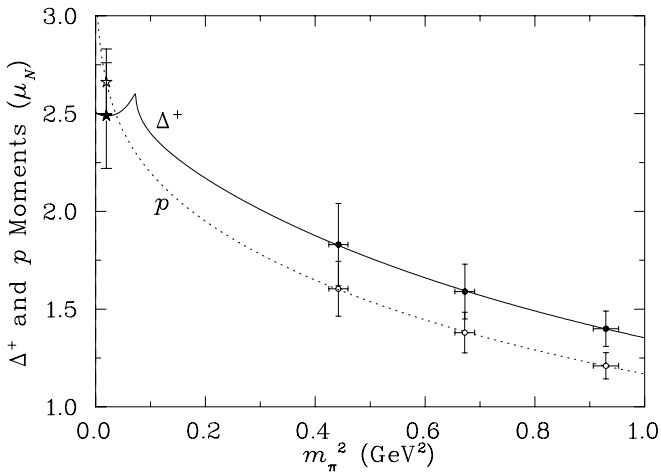, height=\hsize, angle=90}}
\caption{The fit (solid line) to the lattice QCD data (solid circles) for the $\Delta^{+}$ magnetic moment   
using the extrapolation function, Eq.~(\protect\ref{MM}). 
The theoretical prediction for the $\Delta^+$ magnetic 
moment is indicated by a solid star ($\star$).
The initial experimental value for the $\Delta^+$ 
magnetic moment is discussed in the text.
The proton extrapolation (dotted line) of 
the lattice QCD simulation results of Ref.~\cite{Leinweber:1990dv} (open circles)
is  included for reference.}              
\label{fig:d+}
\end{center}
\end{figure}
\begin{figure}[tbp]
\begin{center}
{\epsfig{file=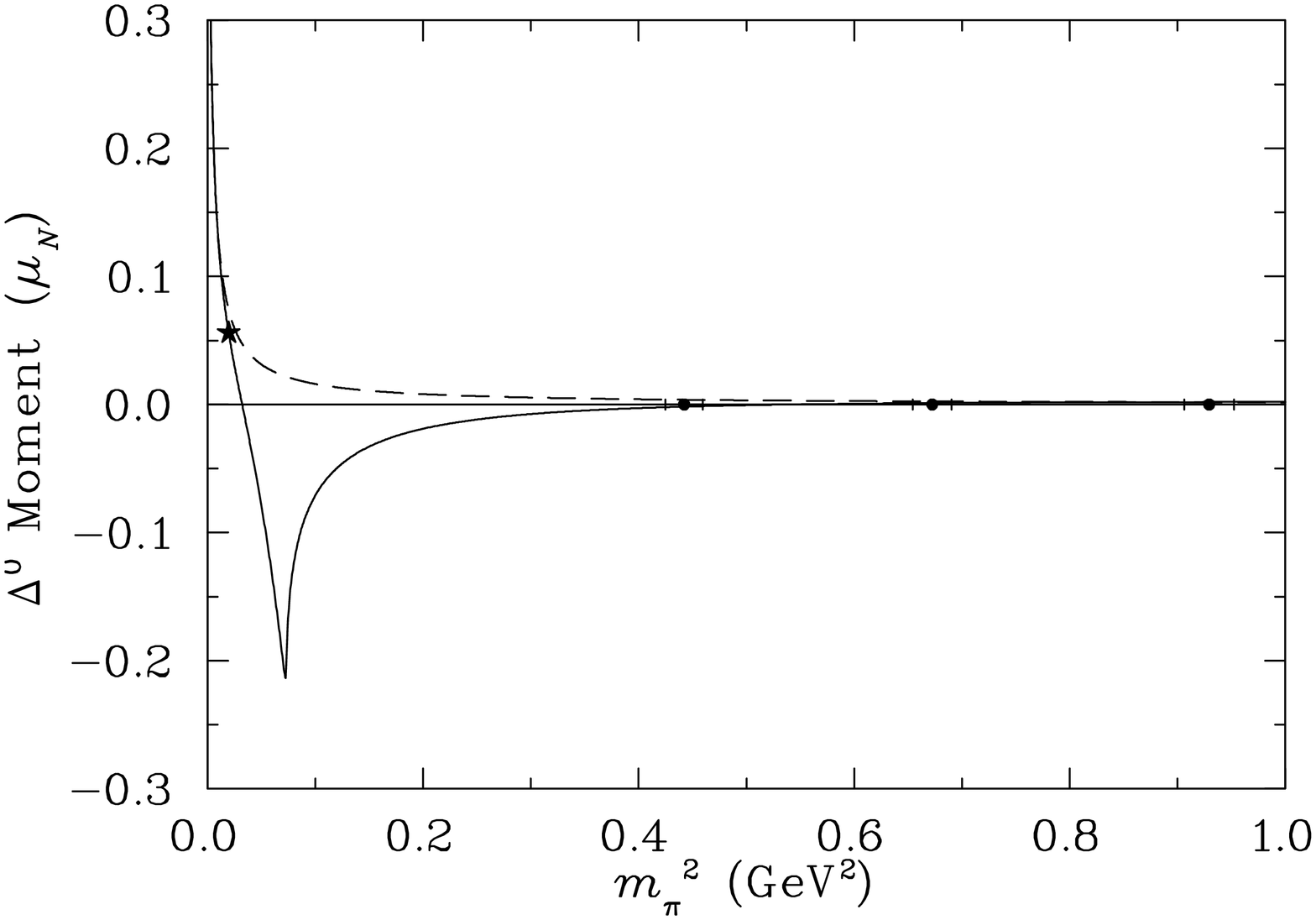, height=\hsize, angle=90}}
\caption{The fit to the lattice QCD data for the magnetic moment of  
the $\Delta^0$, using the extrapolation function, Eq.~(\protect\ref{MM}).
Symbols are as in Fig.~4.
There is currently no experimental value for the $\Delta^0$ magnetic moment.} 
\label{fig:d0}
\end{center}
\end{figure}
\begin{figure}[tbp]
\begin{center}
{\epsfig{file=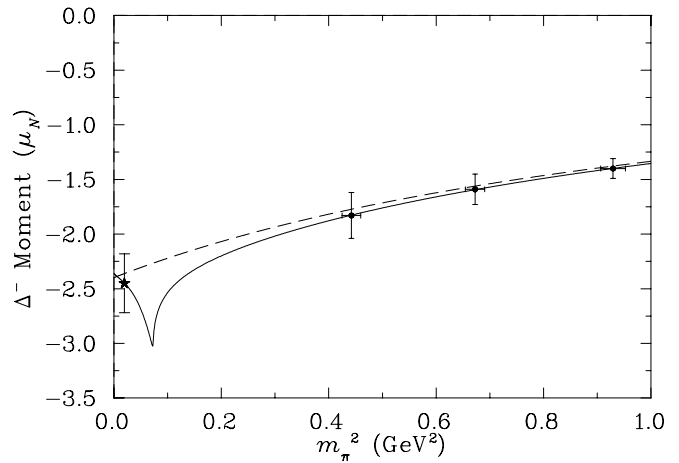, height=\hsize, angle=90}}
\caption{The fit to the lattice QCD data for the magnetic moment of  
the $\Delta^-$, using the extrapolation function, Eq.~(\protect\ref{MM}). 
Symbols are as in Fig.~4.
There is currently
no experimental value for the $\Delta^-$ magnetic moment.}  
\label{fig:d-}
\end{center}
\end{figure}
Therefore decuplet to octet transitions enhance the $\Delta$ magnetic moments 
when the octet decay channel is closed. 
However, as the decay channel opens this physics  
serves to suppress the magnetic moment as the chiral limit is approached.
These transitions are energetically favourable, making them of paramount 
importance in determining the physical properties of $\Delta$ baryons. 

The inclusion of octet--decuplet transitions in octet magnetic 
moment extrapolations is less important.
Significant energy must be borrowed from the vacuum for these transitions to 
occur. The formulation of effective field theory with a finite-range 
regulator naturally suppresses transitions from ground 
state octet baryons to excited state baryons. 
Physically, the finite size of the meson-cloud source suppresses
these high energy contributions.
Hence any new curvature associated 
with octet--decuplet transitions for octet baryons is small \cite{Leinweber:1998ej}. 
Indeed, the inclusion of the $p \to \Delta\, \pi$ channels in the proton moment
extrapolation of Fig.~5 increases the predicted moment by only 0.05\ $\mu_N$ from 2.61
to 2.66\ $\mu_N$.
For this reason octet to decuplet transitions have been omitted in other chiral extrapolation
studies of the octet magnetic moments \cite{Hackett-Jones:2000qk,Cloet:2002eg}.

In the simplest SU(6) quark model with $m_u = m_d$ 
the $\Delta^{+}$ and proton moments 
are equal. 
However most quark models include spin dependent $q-q$ interactions that 
enhance the $\Delta^{+}$ magnetic moment relative to the proton. 
This phenomenology is supported by the lattice results \cite{Leinweber:1992hy}
at large quark masses.
Consequently, previous linearly extrapolated lattice 
predictions \cite{Leinweber:1992hy}, using the 
same lattice data, suggested that the $\Delta^+$ moment should be 
greater than that of the proton. 
We find evidence that supports a different conclusion,
with predicted values for the $\Delta^+$ and proton moments 
lying close at 
2.49(27)\ $\mu_N$ and 2.66(17)\ $\mu_N$, respectively.
The primary reason for this result is the interplay between the three 
different pion loop contributions; 
$\Delta^+ \to \Delta^{++}\, \pi^-$, $n\, \pi^+$ and  $\Delta^0\, \pi^+$.
The transition to  
a $\Delta^{++}\, \pi^-$ state largely cancels the $n\, \pi^+$ contribution,
which dominates the proton magnetic moment. This means that the $\Delta^0\, \pi^+$ transition 
is the main loop contribution to the $\Delta^+$ magnetic moment, where the 
$\Delta^+ \to \Delta^0\, \pi^+$ coupling is weak, relative to that of 
the $p \to n\, \pi^+$ transition.

The proton magnetic moment extrapolation is included 
in Fig.~5, and provides an illustration of the importance of incorporating 
known chiral physics in any extrapolation to the physical world. 
Thus, an experimental
value for the $\Delta^{+}$ magnetic moment would offer very important 
insight into the 
role of spin dependent forces and chiral nonanalytic behaviour 
in the structure of baryon resonances.

\end{section}

\begin{section}{Conclusion}

Finite-range regulated $\chi$PT has been applied to the 
extrapolation of lattice QCD results for 
the decuplet baryon magnetic moments.
The magnetic moments of 
the four $\Delta$ baryons at the resonance pole are determined. 
Experimental values exist only for the 
$\Delta^{++}$ magnetic moment. A result for the $\Delta^{+}$ should be forthcoming
in the near future. The Particle Data Group gives a range 
of 3.7--7.5\ $\mu_N$ for the $\Delta^{++}$ moment, with the two most recent experimental results being 
$\mu_{\Delta^{++}}=4.52 \pm 0.51 \pm 0.45\ \mu_N$ \cite{Bosshard} and 
$\mu_{\Delta^{++}}=6.14 \pm 0.51\ \mu_N$  \cite{LopezCastro:2000ep}. 
Our lattice QCD prediction of $\mu_{\Delta^{++}}=4.99 \pm 0.56\ \mu_N$
compares well with the first of these experimental results.
An interesting result from this investigation 
is the prediction that the $\Delta^+$ magnetic moment lies close to and probably 
below that of the proton moment. 
Therefore a new high precision  experimental measurement of the  $\Delta^+$ moment
would offer valuable insight into spin dependent forces 
and chiral nonanalytic behaviour
of excited states.
The lattice data used in these calculations is now over 10 years old, 
determined with pion masses greater than 600~MeV. 
The arrival of new lattice data 
at lower pion masses is therefore eagerly anticipated, and will help 
constrain the fit parameters and associated statistical uncertainties. 
All of this should be forthcoming 
in the next few years 
and offers an excellent opportunity to test the predictions of QCD.

\end{section}


\begin{acknowledgments}
We thank Ross Young for interesting and helpful discussions.
This work was supported by the Australian Research Council and the University
of Adelaide.
\end{acknowledgments}


\begin{thebibliography}{20}

\bibitem{Hagiwara:pw}
K.~Hagiwara {\it et al.}  [Particle Data Group Collaboration],
Phys.\ Rev.\ D {\bf 66}, 010001 (2002).

\bibitem{Bosshard}
A.~Bosshard {\it et al.},
Phys.\ Rev.\ D {\bf 44}, 1962 (1991).
%

\bibitem{LopezCastro:2000ep}
G.~Lopez Castro and A.~Mariano,
Nucl.\ Phys.\ A {\bf 697}, 440 (2002).

\bibitem{Kotulla}
M.~Kotulla  [TAPS and A2 Collaborations],
{\it Prepared for Hirschegg '01: Structure of Hadrons: 29th International Workshop on Gross 
Properties of Nuclei and Nuclear Excitations, Hirschegg, Austria, 14-20 Jan 2001}.
%

\bibitem{Drechsel:2001qu}
D.~Drechsel and M.~Vanderhaeghen,
Phys.\ Rev.\ C {\bf 64}, 065202 (2001).

\bibitem{Kotulla:2002cg}
M.~Kotulla {\it et al.},
Phys.\ Rev.\ Lett.\  {\bf 89}, 272001 (2002).

\bibitem{Leinweber:1998ej}
D.~B.~Leinweber, D.~H.~Lu and A.~W.~Thomas,
Phys.\ Rev.\ D {\bf 60}, 034014 (1999).

\bibitem{Hackett-Jones:2000qk}
E.~J.~Hackett-Jones, D.~B.~Leinweber and A.~W.~Thomas,
Phys.\ Lett.\ B {\bf 489}, 143 (2000).

\bibitem{Leinweber:2001ui}
D.~B.~Leinweber, A.~W.~Thomas and R.~D.~Young,
Phys.\ Rev.\ Lett.\  {\bf 86}, 5011 (2001).

\bibitem{Cloet:2002eg}
I.~C.~Cloet, D.~B.~Leinweber and A.~W.~Thomas,
Phys.\ Rev.\ C {\bf 65}, 062201 (2002).

%
\bibitem{Cloet}I.~C.~Cloet, D.~B.~Leinweber and A.~W.~Thomas, in Proc. of 
the Joint Workshop on ``Physics at Japanese Hadron Facility'', eds. V. Guzey {\it et al.}, 
125-135, World Scientific (2002),
arXiv:nucl-th/0211027.

\bibitem{Young:2002ib}
R.~D.~Young, D.~B.~Leinweber and A.~W.~Thomas,
{\em To appear in: Progress in Particle and Nuclear Physics}, 
arXiv:hep-lat/0212031.

\bibitem{Donoghue:1998bs}
J.~F.~Donoghue, B.~R.~Holstein and B.~Borasoy,
Phys.\ Rev.\ D {\bf 59}, 036002 (1999).

\bibitem{Banerjee:1995wz}
M.~K.~Banerjee and J.~Milana,
Phys.\ Rev.\ D {\bf 54}, 5804 (1996).

%
\bibitem{Correction} This definition of $\mu_i$ corrects a sign error 
in Ref.~\cite{Banerjee:1995wz}. There the $\beta_j^i\, \ldots$ term is 
preceded by a minus sign.
%

\bibitem{Hatsuda:tt}
T.~Hatsuda,
Phys.\ Rev.\ Lett.\  {\bf 65}, 543 (1990).

\bibitem{SVW}
S. V. Wright, Ph. D. thesis (The University of Adelaide, 2002).

\bibitem{Bernard:2002yk}
C.~Bernard, S.~Hashimoto, D.~B.~Leinweber, P.~Lepage, E.~Pallante, S.~R.~Sharpe and H.~Wittig,
arXiv:hep-lat/0209086.

\bibitem{Thomas:2002sj}
A.~W.~Thomas,
``Chiral extrapolation of hadronic observables'',
arXiv:hep-lat/0208023.


\bibitem{Li:1971vr}
L.~F.~Li and H.~Pagels,
Phys.\ Rev.\ Lett.\  {\bf 26}, 1204 (1971).

\bibitem{Butler:1992pn}
M.~N.~Butler, M.~J.~Savage and R.~P.~Springer,
Nucl.\ Phys.\ B {\bf 399}, 69 (1993).

\bibitem{Jenkins:1992pi}
E.~Jenkins, M.~E.~Luke, A.~V.~Manohar and M.~J.~Savage,
Phys.\ Lett.\ B {\bf 302}, 482 (1993)
[Erratum-ibid.\ B {\bf 388}, 866 (1996)].

\bibitem{Butler}
M.~N.~Butler, M.~J.~Savage and R.~P.~Springer,
Phys.\ Rev.\ D {\bf 49}, 3459 (1994).

\bibitem{ElAmiri:xa}
M.~El Amiri, J.~Pestieau and G.~Lopez Castro,
Nucl.\ Phys.\ A {\bf 543}, 673 (1992).

\bibitem{Guichon:1982zk}
P.~A.~M.~Guichon, G.~A.~Miller and A.~W.~Thomas,
Phys.\ Lett.\ B {\bf 124}, 109 (1983).

\bibitem{Leinweber:1992hy}
D.~B.~Leinweber, T.~Draper and R.~M.~Woloshyn,
Phys.\ Rev.\ D {\bf 46}, 3067 (1992).

%
\bibitem{Comment} Magnetic Moment contributions from the coupling of the electromagnetic 
current directly to sea-quark loops are neglected in these simulations. However, the net
effect from {\em u, d} and {\em s} sea-quark loops is small due to approximate 
SU(3)--flavour symmetry where the net loop contribution vanishes.
%

%
\bibitem{BE} B. Efron, SIAM Rev. {\bf 21}, 460 (1979).
%

\bibitem{Leinweber:1990dv}
D.~B.~Leinweber, R.~M.~Woloshyn and T.~Draper,
Phys.\ Rev.\ D {\bf 43}, 1659 (1991).

\end{thebibliography}
\end{document}